\documentclass[12pt]{iopart}
\usepackage{epsfig,amssymb}
\begin{document}
\title{\bf Free expansion of two-dimensional condensates with a vortex}
\author {O. Ho{\c s}ten$^1$,
P. Vignolo$^2$, A. Minguzzi$^2$\footnote{To whom correspondence should
be addressed (minguzzi@sns.it)},
B. Tanatar$^3$, and M. P. Tosi$^2$}
\address{$^1$Department of Physics Engineering, Hacettepe 
University, Beytepe, 06534 Ankara, Turkey\\
$^2$INFM and Classe di Scienze, Scuola Normale Superiore, I-56126 Pisa,
Italy\\
$^3$Department of Physics, Bilkent University, Bilkent, 06533 Ankara,
Turkey}
\begin{abstract}
We study the free expansion of a pancake-shaped Bose-condensed gas,
which is
initially trapped under harmonic confinement and containing  a
vortex at its centre. In the case of a radial expansion holding fixed
the axial confinement we consider various models for the interactions,  
depending on the thickness of the condensate relative
to the value of the scattering length. We are thus able to
evaluate different scattering regimes ranging from
quasi-three-dimensional (Q3D) to strictly two-dimensional (2D). 
We find that as the system goes from Q3D to 2D the expansion rate of
the condensate increases whereas that
of the vortex core decreases. In the Q3D scattering regime we also
examine a fully free expansion in 3D and 
find oscillatory behaviour for the vortex core 
radius: an initial fast expansion of the vortex core is followed by a
slowing down. Such a nonuniform expansion rate of the vortex 
core may be taken into account in designing new experiments.
\end{abstract}
\maketitle
\newpage

\section{Introduction}

Two-dimensional (2D) condensates in harmonic confinement
are attracting a lot of attention.
 By varying the anisotropy parameter, measured as the 
ratio between the trap frequencies in the $z$- and the planar directions, 
flatter and flatter (pancake-shaped) 
condensates are being produced in magnetic or optical traps \cite{1,2,3} 
with the ultimate goal of 
observing the special features of low dimensionality. As the gas
approaches the 2D limit, its collisional properties start to influence 
the  boson-boson coupling parameter, which becomes dependent on the
density of the system in contrast to the situation in 3D where  at low
density the
interactions are described by a constant coupling strength.

The quantized vortex states are important in establishing the superfluid 
nature 
of the condensates \cite{leggett,4,5,6,7}. Current experiments
on vortices \cite{8,8bis} 
 have utilized elongated geometries, but one may
expect that studies on pancake-shaped condensates will follow.  
Understanding vortices in 2D is
important since they are expected to play a role in the transition
from the superfluid to the normal state \cite{9}. Their
role in inhomogeneous systems such as trapped gases is 
less well known.

From the experimental point of view, 
observing a vortex in a trapped gas is difficult because of the 
smallness of the core region compared to the size of the gaseous cloud. 
Several solutions have been suggested to circumvent this problem.
The size of the core relative to the cloud radius is 
larger in systems made of a smaller number of particles. 
Another possibility is to create vortices with large angular momentum
as achieved in a recent experiment \cite{10}. The vortex core increases
in a free expansion after the 
release of the trap. Indeed, it has been suggested \cite{5}
that the vortex core may expand faster than  
the condensate cloud. Vortices were first
experimentally observed by releasing the trap and allowing the
condensate to expand \cite{8}.

Several theoretical works have studied  the free expansion of 
vortices in Bose condensed systems inside 
anisotropic trap potentials employing various 3D models for the
coupling constant \cite{5,6,9,10}. In a recent work \cite{13} we have 
calculated the equilibrium density profiles
of pancake-shaped condensates using various models for the
coupling  and we have found that, when the condensate is in the 2D
regime for what concerns its collisional properties, its  density profile is
very different in size from what is predicted by the 3D model.

We consider in this paper pancake-shaped condensates 
with a vortex, employing  various models for the coupling parameters
to describe 
the 2D regimes of interest. We study both their equilibrium profile 
and the expansion properties after the trapping potential is released.
By calculating the time evolution of the condensate size and of the vortex core
radius and their expansion rates in 2D we again establish how
the crossover regime may be identified in such experiments. 
We find that the condensate
cloud expands faster in its plane as the system goes from being Q3D to 2D. 
The expansion rate of the vortex core radius, on the other hand,
decreases in the dimensionality crossover.
We also find that the expansion characteristics of the condensate cloud
and of the vortex core 
are rather different when the trap potential is released only in the
radial plane or in both  the $z$- and the perpendicular directions.
In the latter case the velocity of the vortex core can be 
greater than that of the condensate cloud in the initial 
phase of the expansion.

The paper is organized as follows. We first introduce our 
description of condensates with a vortex for different regimes
of scattering properties in Sec.\,II. We then present our 
results in Sec.\,III and conclude with a brief summary in Sec.\,IV.

\section{The model}
\label{model}
We consider a dilute Bose-condensed gas at zero temperature under 
anisotropic harmonic confinement characterized by a trap frequency
$\omega_\perp$ in the $x$-$y$ plane  and an axial trap frequency 
$\omega_z=\lambda\omega_\perp$, with $\lambda\gg 1$.
The motion in the $z$-direction is frozen and  we may describe
the ground state of the condensate  by the
wave function $\psi(r)$ in the $x$-$y$ plane, to be determined from a
2D Schr{\"o}dinger equation. The dynamics of the 2D condensate which
accommodates a quantized 
vortex state is described by a time-dependent nonlinear 
Schr{\"o}dinger equation (NLSE), which reads
\begin{equation}
\fl
i\hbar{\partial\over\partial t}\psi(r,t)=\left[-{\hbar^2\over 2m}
\nabla^2+{\hbar^2\kappa^2\over 2m r^2}+{1\over 2}m\omega_\perp^2r^2+
g(\psi(r,t))|\psi(r,t)|^2\right]\psi(r,t)\, .
\end{equation}
Here, $\hbar\kappa$ is the quantized angular momentum along the 
$z$-axis, $m$ is the atomic mass,  and $g(\psi)$ is the coupling 
strength in 2D which depends
on the condensate wave function as discussed below. We have assumed
that the order parameter entering the NLSE can be written
as $\Phi({\bf r},t)=\psi(r,t)e^{i\kappa\phi}$ where $\phi$ is the
azimuthal angle. We consider repulsive interactions between the
atoms in the condensate, which enter the coupling strength
$g$ through a positive scattering length $a$. 

As we have discussed previously \cite{13}, the density profile of the 
condensate reflects  the modified shape of the confining 
potential and the modified scattering properties. 
When the linear dimension $a_z=(\hbar/m\omega_z)^{1/2}$ 
of the condensate cloud in the $z$-direction
 is much larger than the 3D scattering 
length ($a_z\gg a$), the collisions take place in three dimensions 
and the coupling constant to be used in the NLSE is
\begin{equation}
g_{Q3D}=2\sqrt{2\pi}{\hbar^2\over m}{a\over a_z}\, ,
\end{equation}
including only the geometrical effects of the reduced dimensionality.

When the anisotropy further increases and $a_z$ becomes comparable
with  $a$ ( $a_z \simeq a$),  the collisions start to be influenced by
the confinement  
in the $z$-direction and the interaction strength assumes a form
appropriate to a quasi-two-dimensional condensate,
\begin{equation}
g_{Q2D}={2\sqrt{2\pi}(\hbar^2/m)(a/a_z)\over 1+{a\over \sqrt{2\pi}a_z}
|\ln{(g_{Q2D}n(2\pi m/\hbar^2)a_z^2)}|}\, .
\end{equation}
Here, $n(r,t)=N|\psi(r,t)|^2$ is the areal density of the condensate
cloud, whose dependence on the radial coordinate and on time is
to be calculated self-consistently during the numerical solution of 
Eq.\,(1).
This expression was originally derived by Petrov {\it et al.}~\cite{14,15}
by studying the scattering amplitude in  a system which is harmonically 
confined in the $z$-direction and homogeneous in the $x$-$y$ plane.
 The same result has also been obtained by Lee {\it et al.} \cite{16,17}
within a many-body $T-$matrix approach. Note that the coupling strength
now depends on density,  as is to be expected for 2D collisions.
Furthermore, $g_{Q2D}$ is given by an implicit relation which
has to be solved numerically during the solution of the
NLSE \cite{18}.

Finally, the strictly 2D regime is approached as $a_z$ 
becomes much smaller than $a$ ($a_z\ll a$) and the system is 
described by the coupling
\begin{equation}
g_{2D}={4\pi\hbar^2\over m}\,{1\over |\ln{(na^2)}|}\, ,
\end{equation}
as was first derived by Schick \cite{19} for a homogeneous Bose 
gas of hard disks \cite{notina}. The use of the coupling strength $g_{2D}$ for
inhomogeneous systems,  
involving a dependence 
on the local density, has been proposed by 
Shevchenko \cite{20} and more recently by Kolomeisky 
{\it et al.}~\cite{21}. Note that also $g_{2D}$ has a spatial
dependence due to $n$ but, contrary to $g_{Q2D}$, 
carries no information on the 
confinement in the $z$-direction.

We solve the time-dependent NLSE iteratively by discretization using a 
split-step Crank-Nicholson scheme \cite{22}. In the case of 2D
calculations a simple one-dimensional array of grid points is
sufficient to describe
the spatial part of $\psi(r,t)$ at each time step. The 
3D calculations that we subsequently report for $\psi(r,z,t)$
as a solution of the standard Gross-Pitaevski equation (GPE)
involve a two-dimensional grid 
because of the cylindrical symmetry of this wave function.

\section{Results and discussion}

\subsection{Equilibrium profiles and radial expansion}

The numerical solution of the  NLSE with a 
centrifugal term in equilibrium conditions 
gives the ground-state  wave function  of the condensate with a 
single vortex ($\kappa=1$).
We first take values of the anisotropy parameter and of the    scattering length
as appropriate for $^{23}$Na atoms in the experiment of G{\"o}rlitz
{\it et al.}~\cite{1} ($\lambda=26.33$ and  $a=2.8$\,nm, so that 
$a/a_z=3.8\times 10^{-3}$). We scale the radial coordinate by the 
harmonic oscillator length
$a_\perp=(\hbar/m\omega_\perp)^{1/2}$ and the radial wave function
by $1/a_\perp$.  We take $N=5000$
since it is easier to observe vortices for smaller numbers of
particles. For these parameters, the system undergoes
collisions in 3D 
but has 2D characteristics for what concerns the confinement effects.
As in the absence of the vortex, we find that the Q3D and Q2D models yield
wave function profiles identical to each other as shown in Fig.\,1(a),
whereas the 2D model produces a quantitatively very different result.

We next look at the planar free expansion of the condensate
with a vortex for the same parameters, holding fixed the confinement
in the axial direction. The time-dependent NLSE
is solved after the trap potential ${1\over 2}m\omega_\perp^2 r^2$ is
switched off at time $t=0$.  Figure 1(b) shows the time dependence of the 
root-mean-square (rms) value of the radial coordinate 
$r_{rms}=\langle r^2\rangle^{1/2}$, which 
describes the size of the condensate cloud,
and of the vortex core radius $R_c$ in three different models of the 
coupling. As in \cite{6}, the core radius is chosen to be  the value 
of $r$ where the condensate density reaches $1/e$ of its peak value, viz. 
$|\psi(R_c)|^2= e^{-1}\hbox{max}\{|\psi(r)|^2\}$. 
We see from Fig.\,1(b) that the strictly 2D model predicts
a much faster expansion compared to the Q3D and Q2D models, which 
again yield identical results for the present set of parameters. The
vortex core 
expands much more slowly than the condensate cloud, as indicated in 
the same figure. We also  show in Fig.\,1(c) the velocity of expansion
of 
the atomic cloud and of the vortex core as functions of time.
After an initial accelerated motion both velocities attain a constant 
value for $t\omega_\perp\gtrsim 2$.

We increase the anisotropy parameter to $\lambda=2\times 10^5$ and
take $a/a_z=0.33$ in order to investigate the regime of crossover from
3D to 2D.  
As shown in Fig.\,2(a), the profiles of the  wave function with a vortex all 
look similar in this regime of parameters. 
Consequently the time dependence of the radial coordinate and of the
vortex core radius display similar behaviours during the expansion, as
shown in Fig.\,2(b).
We note that the expansion of the condensate cloud is faster 
than in the Q3D regime. This is also indicated 
by the higher velocities depicted in Fig.\,2(c). The velocity of the
vortex core radius, however, decreases compared to the previous regime. 
The small wiggles seen in $\dot{R}_c$ in the inset of 
Fig.\,2(c) are due to numerical differentiation of the results shown in Fig.\,2(b) and are not expected to be physical.

The system enters the strictly 2D regime 
when we further increase the scattering length to $a/a_z=2.68$.
 In this case, the vortex
wave function is best described by the 2D model using $g_{2D}$, although
the results of the Q2D model using $g_{Q2D}$ are very similar.
As the initial shape of the cloud in the Q3D model extends out to a
larger radial  
distance  as shown in Fig.\,3(a), one may again
distinctly identify the truly 2D regime by the wave function profile.
Furthermore, we observe in Fig.\,3(b) that the atomic cloud expands 
slightly faster compared to the previous Q3D and crossover regimes.
This is also evident in Fig.\,3(c), where the expansion velocity of 
the cloud is larger. The velocity $\dot{R}_c$ of the vortex core radius,
on the other hand, is very small in the strictly 2D regime.

The above results indicate that the initially trapped Bose condensate
expands faster in the $x$-$y$ plane as the system moves from being 3D
to 2D. 
The opposite behaviour is observed for the vortex core. However, the 
expansion rate of the condensate cloud remains larger than that of the 
vortex in all cases that we have considered. Our numerical results 
are in good agreement with previous studies of vortices in
Bose-condensed fluids \cite{6,8} for similar 
values of the parameters.

\subsection{Three-dimensional expansion}

We focus now on the case of a free expansion in  3D space, adopting 
the set of experimentally relevant parameters 
as in Fig.~1. In this case the presence of the confinement
does not affect the binary collisions between the atoms and the system is well
described by the 3D time-dependent 
GPE  with coupling constant
$g_{3D}=4\pi\hbar^2 
a/m$.

We have first tested the consistence of our Q3D model by solving the
GPE for a 3D anisotropic system to
study the expansion when only the radial trap potential 
$m\omega_\perp^2 r^2/2$ is turned off ({\it i.e.} the potential 
$m\omega_z^2 z^2/2$ remains throughout the expansion). The results 
are identical to those obtained earlier in Figure 1 with the 2D kinematics and 
the Q3D scattering properties, showing  the correctness of our 
physical picture.

We have then studied the expansion properties of a 3D 
anisotropic condensate when the trap potentials in both 
$r$ and $z$ directions are released. Essentially identical results are
obtained  for two different choices of the initial configuration, {\it
i.e.} starting from the ground state of the 3D GPE or from the ground
state of the 2D GPE 
with Q3D coupling combined with a Gaussian profile in the third
direction. In both cases we calculate the time-dependent
behaviour of the rms values of the radial and axial coordinates, as
well as that of 
the core radius $R_c$.
The results are shown in Fig.~4. We observe a
much faster expansion 
for the $z$ coordinate, reflecting  the initially
tighter axial confinement. It also appears that the vortex core 
radius undergoes a slow oscillation, as is
more evident in Fig.\,4(b) where we plot the velocities.

The oscillatory behaviour of the vortex core can be understood as
 being closely related  to the repulsive interactions, which strongly
 affect the 
density profile. As compared  to the 
ground state of the noninteracting case, the repulsive interactions give
 rise to a smaller vortex 
core radius and to a larger rms radius.  
 The density suddenly
decreases when the trap is released in both $r$ and $z$ directions,
as a result of the fast expansion in the
$z$ direction, and this causes the nonlinear interaction energy to
 rapidly vanish on the time scale of the radial
 expansion.
Therefore, in the first part of the expansion (for $t\omega_\perp\lesssim 0.8$
in our study) the compressed density at  the centre of the cloud 
 expands faster than the remaining part, with the result that
the core expansion rate is larger than that of the condensate cloud. Similar
 results were also found by Dalfovo and Modugno \cite{6}. 
Then, when the density is so low that the interaction energy becomes
 negligible, 
the vortex cannot continue to expand faster than the cloud
and the velocity $\dot{R}_c$ even reverses sign around
 $t\omega_\perp\approx 1.5$, as shown in Fig.\,4(b). 

We have also studied the expansion properties of systems with 
$N=5\times 10^4$ and $N=5\times 10^5$ particles and the same trap 
frequencies and scattering length as in Fig.~1, 
finding again  a much rapid expansion in the $z$ direction and 
oscillatory behaviour in the vortex core radius. These results are not
shown since they  are qualitatively very  similar to those in Fig.~4.

\section{Summary and concluding remarks}
In the perspective  of experimental investigations on low-dimensional
condensates, we have studied the free expansion of pancake-shaped Bose-Einstein
condensates with a vortex. By choosing various values of the trap 
anisotropy and scattering length we have explored different scattering 
regimes, from a quasi-three-dimensional regime to a strictly two-dimensional
one. We have described the cloud at mean field level, taking 
as coupling strengths for the boson-boson interactions
the density-dependent expressions derived by Petrov {\it et al.} \cite{14}
for the quasi-2D regime and by Kolomeisky {\it et al.} \cite{21}
for the strictly 2D regime. 

We have considered both the case of a 2D expansion keeping
the confinement along $z$ fixed and the case of a fully 3D expansion.
In the 2D case we have observed that with increasing anisotropy
the expansion rate of the cloud increases while that of the vortex
core decreases. 
In the 3D expansion, performed in the regime where the collisions are 
three-dimensional and the Q3D model is the most appropriate,
 the expansion rate of the vortex core exhibits an
oscillatory behaviour due to the interplay between nonlinear 
interactions and anisotropic confinement. 

Whereas in this work we have restricted our analysis to the case
of a condensate with a single vortex at $T=0$, it may also be
interesting to investigate the role of the
noncondensate cloud at finite temperature, as 
 vortices are intimately connected
with the Kosterlitz-Thouless phase transition. To 
explore this area it would be necessary to take into account
multiple vortices within the condensate cloud \cite{7}.

\ack
We acknowledge partial support by INFM within the PRA-Photonmatter Initiative.
O.\,H. thanks TUBITAK-BAYG. B.\,T. acknowledges the support of TUBITAK,
NATO-SfP, MSB-KOBRA and  TUBA, and the hospitality of Scuola Normale
Superiore during part of this work.

\newpage

\begin{figure}
\epsfig{file=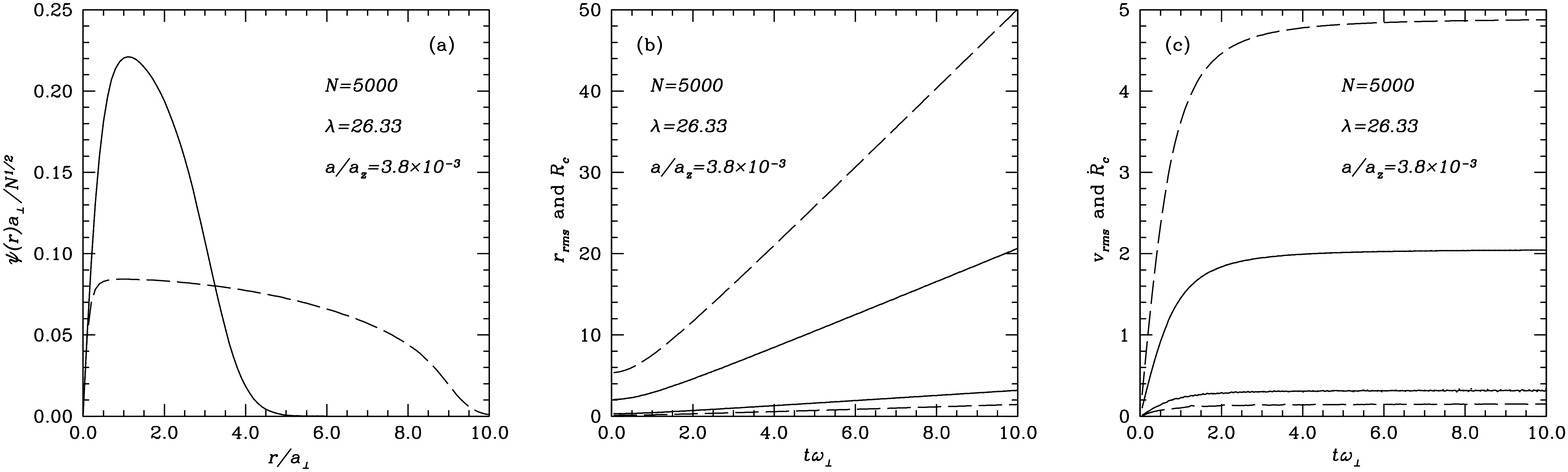,width=1\linewidth}
\caption{Properties of a condensate with a vortex within the
models described  by $g_{Q3D}$ (dotted lines), $g_{Q2D}$ (solid lines)
and $g_{2D}$ (dashed 
lines), for $N=5000$, $\lambda=26.33$  and $a/a_z=3.8\times
10^{-3}$. (a) The vortex-state equilibrium wave function.
(b) The time dependence of the root-mean-square (rms) value of the
radial coordinate $r_{rms}$ (upper curves) and that of the vortex core
radius $R_c$ (lower curves). (c) The time
dependence of the rms value $v_{rms}$ of the velocity of the radial coordinate
 (upper curves) and that of the velocity $\dot{R}_c$  of the vortex core
radius (lower curves).
The dotted lines are superimposed to the solid lines for this set
of parameters.}
\end{figure}

\begin{figure}
\epsfig{file=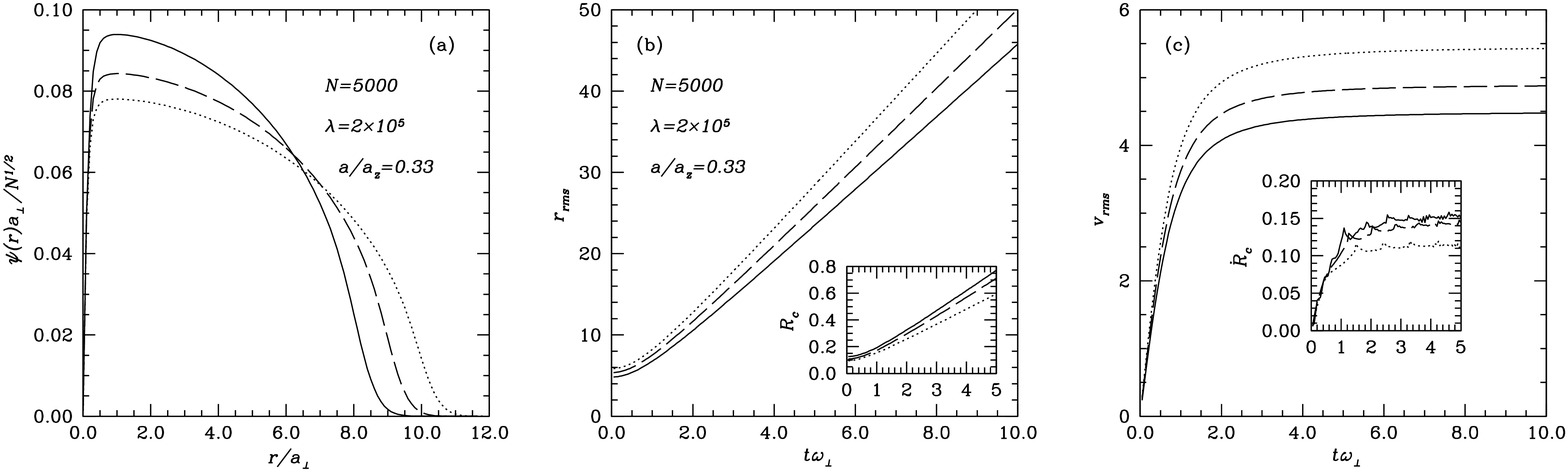,width=1\linewidth}
\caption{Properties of a condensate with a vortex
 within the models described
by $g_{Q3D}$ (dotted lines), $g_{Q2D}$ (solid lines) and $g_{2D}$ (dashed
lines), for $N=5000$, $\lambda=2\times 10^5$ and $a/a_z=0.33$. 
(a) The vortex-state wave function at equilibrium.
(b) The time dependence of rms radial coordinate $r_{rms}$ and of the 
vortex core radius $R_c$ (inset). 
(c) The time dependence of the rms velocity $v_{rms}$ of the radial 
coordinate  and of the velocity $\dot{R}_c$ of the 
vortex core radius (inset).}
\end{figure}

\begin{figure}
\epsfig{file=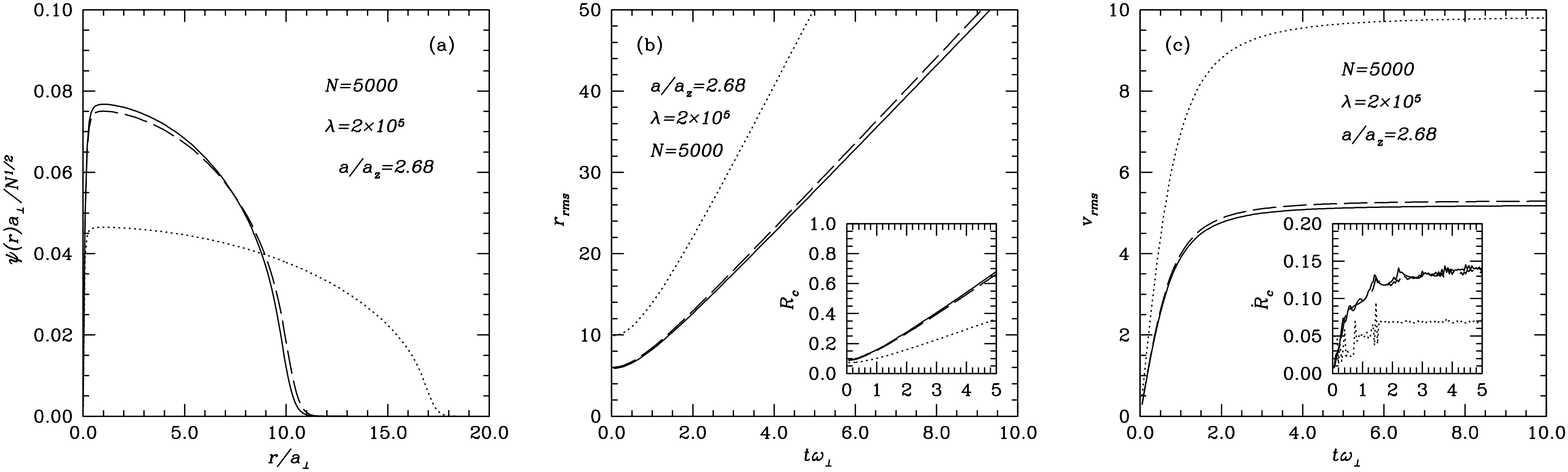,width=1\linewidth}
\caption{Properties of a condensate with a vortex
 within the models described
by $g_{Q3D}$ (dotted lines), $g_{Q2D}$ (solid lines) and $g_{2D}$ (dashed
lines), for $N=5000$, $\lambda=2\times 10^5$ and $a/a_z=2.68$.
(a) The vortex-state wave function at equilibrium.
(b) The time dependence of the rms radial coordinate $r_{rms}$ and of the 
vortex core radius $R_c$ (inset). 
(c) The time dependence of the rms velocity $v_{rms}$ of the radial 
coordinate  and of the velocity $\dot{R}_c$ of the 
vortex core radius  (inset).}
\end{figure}

\begin{figure}
\epsfig{file=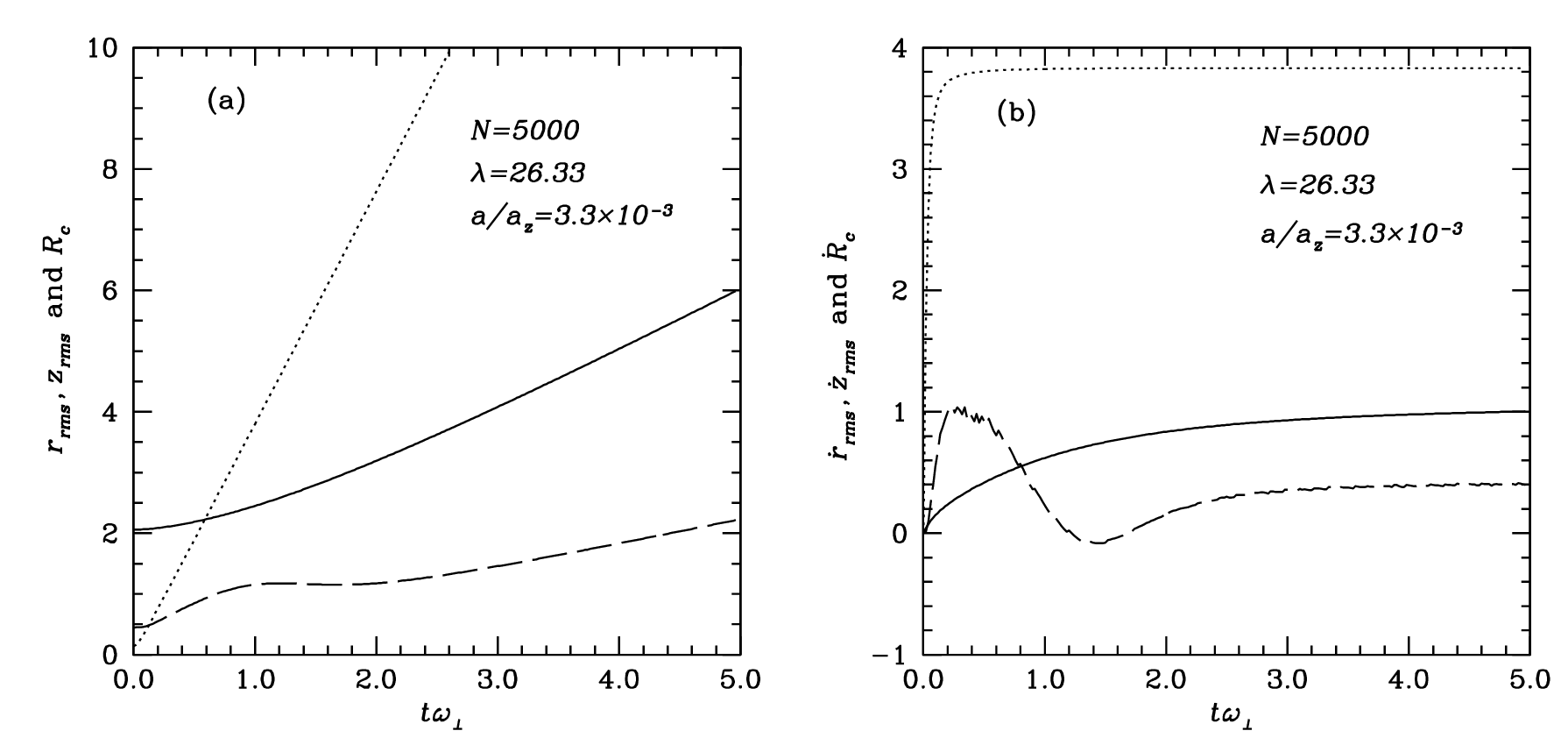,width=1\linewidth}
\caption{(a) The time dependence of the rms
values of the radial (solid lines) and $z$-coordinates (dotted lines), 
and that of the vortex core radius (dashed lines), for a 3D system with 
$N=5000$, $\lambda=26.33$ and $a/a_z=3.8\times 10^{-3}$.
(b) The time dependence of the velocities $\dot{r}_{rms}$,
$\dot{z}_{rms}$ and $\dot{R}_c$ for the same system.}
\end{figure}

\begin{thebibliography}{10}
\bibitem{18} In our previous work (Ref.\,13) we have
approximated the expression given in Eq.\,(3) by 
$g_{Q2D}={2\sqrt{2\pi}(\hbar^2/m)(a/a_z)/[1+(a/\sqrt{2\pi}a_z)
|\ln{(2(2\pi)^{3/2}n a a_z)}|}]$,
which amounts to approximating $g_{Q2D}$  inside
the logarithm in Eq.~(3) by its zeroth order value
$2\sqrt{2\pi}(\hbar^2/m)a/a_z$. We have checked that by using
Eq.\,(3) the main results
of Ref.\,13 
are not altered to any significant extent. Indeed, the implicit
expression for $g_{Q2D}$ can
change the wave function profiles only  for extremely large values
of the anisotropy parameter 
$\lambda$.
\bibitem{notina}
In the definition of  $g_{2D}$  we are using  the 3D
scattering length $a$. The work of Petrov and Shlyapnikov \cite{15}
and of Lee and Morgan \cite{17} indicates that the use of the 2D
scattering length 
$a_{2D}$ would be more appropriate, but since for a
 2D system $a_{2D}=ae^{\gamma}$
(where $\gamma\simeq 0.577$ is the Euler-Mascheroni constant), in the
dilute-gas  regime ($na^2\ll 1$) the difference between the two
models goes beyond the accuracy of the mean-field description.
\end{thebibliography}
\end{document}